\documentstyle{article}

\newcommand{\beq}[1]{\begin{equation} \label{#1} }
\newcommand{\eeq}   {\end{equation}}
\newcommand{\ds}{\displaystyle \mathstrut}
\newcommand{\Frac}[2]{\frac{\textstyle \mathstrut #1}
{\textstyle \mathstrut #2}}
\newcommand{\ve}{\varepsilon}
\newcommand{\valpha}{\mbox{\boldmath $\alpha$}}
\newcommand{\vsigma}{\mbox{\boldmath $\sigma$}}
\newcommand{\ov}[1]{\overline{ #1}}

\begin{document}
\large

\title{INDUCED PSEUDOSCALAR COUPLING IN MUON CAPTURE AND
SECOND-ORDER CORRECTIONS}
\author{A.L.Barabanov\\
{\it RRC "Kurchatov Institute", Moscow 123182, Russia}}
\date{\mbox{}}
\maketitle

\begin{abstract}
The hyperfine effect in muon capture rate is analyzed with taking
into account of the second-order terms in $1/M$ in the
non-relativistic Hamiltonian. It is shown that in the situation of
the dominance of the squared first-order terms the interference of
zero-order and second-order terms has no significant influence.
General expression for neutrino (recoil nucleus) angular
distribution in muon capture by nucleus with non-zero spin is
proposed. High sensitivity to $g_P$ of the term related with
alignment of the initial mesic atom is discussed in the
approximation of dominating Gamow--Teller matrix element.
\end{abstract}

\section{Introduction}

PCAC hypothesis gives a relation between the induced pseudoscalar
form factor and the axial vector one. In muon capture one expects
$g_P\simeq 7g_A\simeq -8.4$. This prediction is in agreement with
measurements of $g_P$ on proton and some light nuclei as $^{12}$C
and $^{23}$Na (see, e.g., \cite{Joh96}). However, the experiments
on muon capture by $^6$Li \cite{Deu68} and $^{28}$Si \cite{Bru95}
yielded the values for $g_P$ close to zero. To clarify a situation
with this suppression that partly may be due to medium effects
further investigations are of interest.

One of the possible ways to measure $g_P$ is the hyperfine effect
(see, e.g., \cite{Joh96,Wia97}). A ratio of muon capture rates
$\Lambda_+/\Lambda_-$ from different hyperfine sublevels is rather
sensitive to the induced pseudoscalar coupling. In particular, if
Gamow--Teller matrix element dominates in the transition
$|J\pi\rangle\to |J-1\,\pi\rangle$, then the ratio
\beq{1.1}
\frac{\Lambda_+}{\Lambda_-}=
\frac{G_P^2}
{\Frac{3(2J+1)}{J}\left(G_A-\Frac{1}{3}G_P\right)^2+
\Frac{J-1}{3J}G_P^2},
\eeq
exhibits the high sensitivity to
$G_P=(E_{\nu}/2M)(g_P-g_A-g_V-g_M)$ or to the form factor $g_P$.
Here $E_{\nu}$ is the neutrino energy, and $M$ is the nucleon
mass. However, it should be noted that the ratio
$\Lambda_+/\Lambda_-$ is proportional to $(1/M)^2$, whereas the
usually used description of the hyperfine effect
\cite{Bal67}-\cite{Wal76} is based on the non-relativistic
Hamiltonian which is of first-order in $1/M$. Thus the analysis
based on the second-order Hamiltonian seems necessary for
consistent description of the hyperfine effect. Such analysis was
the first aim of this work.

The second aim was to obtain the general expression for the
angular distribution of neutrinos (recoil nuclei) in muon capture
by target with non-zero spin $J$. Due to the hyperfine splitting
in the initial mesic atom interference of states with different
angular momenta $F_{\pm}=J\pm 1/2$ gives no contribution to the
differential probability of muon capture. Therefore, the angular
distribution of neutrinos is determined by sum
\beq{1.2}
\frac{dw({\bf n}_{\nu})}{d\Omega}=
\sum_FP(F)\frac{dw^F({\bf n}_{\nu})}{d\Omega},
\eeq
where $P(F)$ is the population of the state $|F\rangle$. Thus,
irrespective of whether the process of capture from states with
different $F$ are detected experimentally or not, the angular
distribution of neutrinos must be calculated separately for each
state of the hyperfine structure.

Explicit expressions for the asymmetry of neutrino emission along
and opposite the direction ${\bf n}_{\mu}$ of mesic atom
polarization were obtained in
Refs.\cite{Pri59},\cite{Bal67},\cite{Gal68}. However, the ensemble
of the initial mesic atoms with spin $F>1/2$ may not only be
polarized but aligned as well. This leads to the anisotropy of
neutrino emission in the directions parallel and orthogonal to the
vector ${\bf n}_{\mu}$. The alignment effect was analyzed for
transitions
$|1/2\,\pi\rangle\to |1/2\,\pi\rangle$ and $|1\,\pi\rangle\to
|0\,\pi\rangle$ in a model-independent (elementary-particle)
approach in Refs.\cite{Hwa78,Con93} and for transitions
$|J\pi\rangle\to |J\pm 1\,\pi\rangle$ in the approximation of
dominating Gamow--Teller matrix element
in Ref.\cite{Bar96}. It was pointed out that the alignment effect
is linear in $1/M$ and is very sensitive to $g_P$. Nevertheless,
the contributions of non-leading matrix elements, in particular,
of velocity terms, are evidently of great interest.

\section{Second-order Hamiltonian}

In the initial 1s-state of mesic atom a captured muon is described
by the 4-component wave function. It can be written in the
non-relativistic approximation as ($\hbar=c=1$)
\beq{2.1}
\psi_{\mu}(\sigma_{\mu},{\bf r}_{\mu},t)=
\psi_{1s}(r_{\mu})
\left(\begin{array}{c}
\varphi_{\mu}(\sigma_{\mu})\\
0 \end{array}\right)e^{-iE_{\mu}t},
\eeq
where $E_{\mu}$ is the total muon energy, including its binding
energy in the atom, $\varphi_{\mu}(\sigma_{\mu})$ is the
two-component spinor, and $\sigma_{\mu}$ is the projection of the
spin $s_{\mu}=1/2$ onto the $z$ axis. A final neutrino with
momentum ${\bf k}_{\nu}$ and projection $\sigma_{\nu}$ of the spin
$s_{\nu}=1/2$ onto the $z$ axis is described by the wave function
\beq{2.2}
\psi_{\nu}(\sigma_{\nu},{\bf r}_{\nu},t)=
u_{\nu}({\bf k}_{\nu},\sigma_{\nu})
e^{i({\bf k}_{\nu}{\bf r}_{\nu}-E_{\nu}t)},
\eeq
where $u_{\nu}({\bf k}_{\nu},\sigma_{\nu})$ is a 4-spinor, and
$E_{\nu}=k_{\nu}$ is the neutrino energy. Assuming that the weak
nucleon-lepton interaction is pointlike, we introduce the lepton
current acting on the nucleon
\beq{2.3}
j_{\lambda}(\sigma_{\mu},\sigma_{\nu})
e^{-i({\bf k}_{\nu}{\bf r}+\omega t)}=
i\psi^+_{\nu}(\sigma_{\nu},{\bf r},t)
\gamma_4(1+\gamma_5)
\psi_{\mu}(\sigma_{\mu},{\bf r},t),
\eeq
where $\omega=E_{\mu}-E_{\nu}$.
Thus the relativistic Hamiltonian for the nucleon into a nucleus
involving in capture of muon being in the spin state
$|\sigma_{\mu}\rangle$ with emission of neutrino in the state
$|{\bf k}_{\nu},\sigma_{\nu}\rangle$ is
\beq{2.4}
\hat H=M\beta+\valpha\hat{\bf p}+U(r)+\hat H_W,
\eeq
\beq{2.5}
\hat H_W=
\frac{G\cos\theta_C}{\sqrt{2}}
j_{\lambda}(\sigma_{\mu},\sigma_{\nu})
e^{-i({\bf k}_{\nu}{\bf r}_{\nu}+\omega t)}
(i\Gamma_{\lambda})\hat\tau_-,
\eeq
where $G$ is the weak-interaction
coupling constant, $\theta_C$ is the Cabibbo angle. The lowering
operator $\hat\tau_-$ acts in isospin space and transforms a
proton into a neutron. For simplicity we take the nucleon-nucleus
potential $U(r)$ in a central form. The operator of the weak
nucleon current is given by
\beq{2.6}
\Gamma_{\lambda}=\gamma_4
\left(g_V\gamma_{\lambda}+
\frac{g_M}{2M}\sigma_{\lambda\rho}k_{\rho}-
g_A\gamma_{\lambda}\gamma_5-
i\frac{g_P}{m}k_{\lambda}\gamma_5\right).
\eeq
It involves the matrices $\sigma_{\lambda\rho}=
(\gamma_{\lambda}\gamma_{\rho}-\gamma_{\rho}\gamma_{\lambda})/2i$,
the muon mass $m$, and the 4-momentum transfer
\beq{2.7}
k_{\lambda}=\nu_{\lambda}-\mu_{\lambda}=
({\bf k}_{\nu},-i\omega),
\eeq
where $\nu_{\lambda}$ and $\mu_{\lambda}$ are the 4-momenta of the
muon and neutrino, respectively. The form factors of vector
interaction $g_V$, axial-vector interaction $g_A$, weak magnetism
$g_M$, and induced pseudoscalar interaction $g_P$ depend on
$k^2=k_{\lambda}k_{\lambda}$.

To go to the non-relativistic description of intranuclear
nucleon we perform the Foldy--Wouthuysen transformation of
relativistic Hamiltonian. Keeping the terms up to second-order in
$1/M$ we obtain for non-relativistic Hamiltonian for $j$-th
nucleon the following expression
\beq{2.8}
\hat H=M+\frac{\hat{\bf p}_j^2}{2M}+U(r_j)+
\frac{\Delta U(r_j)}{8M^2}+
\frac{U'(r_j)}{4M}
(\vsigma[{\bf n_j}\times\frac{\hat{\bf p}_j}{M}])+
\hat h_j(\sigma_{\mu},\sigma_{\nu})e^{-i\omega t},
\eeq
\beq{2.9}
\begin{array}{l}
\hat h_j(\sigma_{\mu},\sigma_{\nu})=
\Frac{G\cos\theta_C}{\sqrt{2}}e^{-i{\bf k}_{\nu}{\bf r}_j}
\left\{ij_4(\sigma_{\mu},\sigma_{\nu})
\left[\phantom{\lefteqn{\Frac{U'}{M}}}
G'_V+G'_P(\vsigma_j{\bf n}_{\nu})+{}\right.\right.
\\[\bigskipamount]
\left.{\enskip}+g'_A(\vsigma_j\Frac{\hat{\bf p}_j}{M})+
iG_1({\bf n}_{\nu}[\vsigma_j\times\Frac{\hat{\bf p}_j}{M}])+
G_2({\bf n}_{\nu}\Frac{\hat{\bf p}_j}{M})+
ig_P\Frac{U'(r_j)}{4M^2}(\vsigma_j{\bf n}_j)\right]+{}
\\[\bigskipamount]
\phantom{\hat h_j(\sigma_{\mu},\sigma_{\nu})=G\;}+
{\bf j}(\sigma_{\mu},\sigma_{\nu})\left[
G'_A\vsigma_j+g'_V\Frac{\hat{\bf p}_j}{M}+
ig_1[\vsigma_j\times\Frac{\hat{\bf p}_j}{M}]+{}\right.
\\[\bigskipamount]
\left.\left.{\enskip}+g_2\left[[\vsigma\times\Frac{\hat{\bf
p}_j}{M}]
\times\Frac{\hat{\bf p}_j}{M}\right]+
g_3\vsigma_j({\bf n}_{\nu}\Frac{\hat{\bf p}_j}{M})+
g_V\Frac{U'(r_j)}{4M^2}[\vsigma_j\times{\bf n}_j]\right]
\right\}\hat\tau_-(j),
\end{array}
\eeq
where ${\bf n}_{\nu}={\bf k}_{\nu}/k_{\nu}$,
${\bf n}_j={\bf r}_j/r_j$, and
\beq{2.10}
\begin{array}{lll}
\lefteqn{G'_V=g_V\left(1+\ve-\Frac{m}{4M}\ve\right)-
g_M\Frac{m}{2M}\ve,} & &
\\[\bigskipamount]
\lefteqn{G'_A=g_A\left(1-\Frac{\ve^2}{2}\right)-
\ve\left(g_V(1-\Frac{\eta}{2})+g_M(1-\eta)\right),} & &
\\[\bigskipamount]
\lefteqn{G'_P=\ve\left((g_P-g_A-g_V)(1-\Frac{\eta}{2})-
g_M(1-\eta)\right),} & &
\\[\bigskipamount]
g'_V=g_V-\Frac{\ve}{2}g_A, &
\lefteqn{g'_A=g_A(1+\Frac{\ve}{2})+g_P\Frac{\eta}{2},} &
\\[\bigskipamount]
G_1=\Frac{\ve}{2}(g_V+g_A+2g_M), &
\lefteqn{G_2=-\Frac{\ve}{2}g_A,} &
\\[\bigskipamount]
g_1=\Frac{1}{2}(\ve g_A-\eta(g_V+2g_M)),\qquad &
g_2=\Frac{1}{2}g_A,\qquad &
g_3=-\Frac{\ve}{2}g_A,
\end{array}
\eeq
\beq{2.11}
\ve=\Frac{E_{\nu}}{2M},\qquad\eta=\Frac{\omega}{2M}.
\eeq
The non-relativistic Hamiltonian of muon capture by free nucleon
(i.e. \mbox{$U(r)=0$}) with second-order corrections was earlier
obtained in Ref.\cite{Fri66}\footnote{Our result differs by the
factor 1/4 in the term $m/(4M)\ve$ in the definition of $G'_V$ and
by the sign of~$g_3$.} as well as in the other forms in
Refs.\cite{Oht66,Con93}.

The zero-order and first-order terms are proportional to the
operators $\hat 1$, $\vsigma_j$,
$\hat{\bf p}_j$, and $(\vsigma_j\hat{\bf p}_j)$. The
second-order terms involve the additional set of operators:
$[\vsigma_j\times\hat{\bf p}_j]$,
$[[\vsigma_j\times\hat{\bf p}_j]\times\hat{\bf p}_j]$,
$\vsigma_j({\bf n}_{\nu}\hat{\bf p}_j)$,
$(\vsigma_j{\bf n}_j)$, and
$[\vsigma_j\times {\bf n}_j]$.

\section{Neutrino angular distribution}

Let $|J_fM_f\rangle$ be the wave function that describes the
internal state of the final nucleus with spin $J_f$ and its
projection $M_f$ onto the $z$ axis. At the same time, the initial
state of the mesic atom with total angular momentum $F$ involving
a nucleus with spin $J_i$ and a muon is represented as
\beq{3.1}
|F\rangle=\sum_{\xi}a_{\xi}(F)
\sum_{M_i\sigma_{\mu}}C^{F\xi}_{J_iM_is_{\mu}\sigma_{\mu}}
|J_iM_i\rangle\psi_{\mu}(\sigma_{\mu}).
\eeq
Polarization and alignment of the ensemble of the mesic atoms with
given total angular momentum $F$ are described by density matrix
\beq{3.2}
\rho_{\xi\xi'}(F)=\ov{a_{\xi}(F)a_{\xi'}^*(F)},\qquad
\sum_{\xi}\rho_{\xi\xi}(F)=1,
\eeq
or by spin-tensors
\beq{3.3}
\tau_{Qq}(F)=\sum_{\xi\xi'}
C^{F\xi'}_{F\xi Qq}\rho_{\xi\xi'}(F),\qquad
\tau_{00}(F)=1.
\eeq
The energy of the neutrino emitted in the fixed transition
$|J_i\rangle\to |J_f\rangle$ is
\beq{3.4}
E_{\nu}=M_f\left[\left(1+
\frac{2Q_{\mu}}{M_f}\right)^{1/2}-1\right]\simeq
Q_{\mu}\left(1-\frac{Q_{\mu}}{2M_f}+\ldots\right),
\eeq
where $Q_{\mu}=M_i+E_{\mu}-M_f$, $M_i$ and $M_f$ are the masses of
the initial and final nuclei, respectively. Using the Fermi rule,
we find that the differential probability of muon capture per unit
time from a given state $|F\rangle$ of the hyperfine structure has
the form
\beq{3.5}
\begin{array}{l}
\Frac{dw^F({\bf n}_{\nu})}{d\Omega}=
\Frac{1}{(2\pi)^2}\times{}
\\[\bigskipamount]
{}\times{\ds\sum\limits_{\sigma_{\nu}M_f}}
\left|{\ds\sum\limits_{\xi}}a_{\xi}(F)
{\ds\sum\limits_{M_i\sigma_{\mu}}}
C^{F\xi}_{J_iM_is_{\mu}\sigma_{\mu}}
\langle J_fM_f|{\ds\sum\limits_{j=1}^A}
\hat h_j(\sigma_{\mu},\sigma_{\nu})|
J_iM_i\rangle\right|^2
\Frac{E_{\nu}^2}{1+E_{\nu}/M_f}.
\end{array}
\eeq

It is convenient to introduce the multipole expansions for the
matrix elements of all operators appearing in the non-relativistic
Hamiltonian. Generalizing the definitions of Ref.\cite{Bal78} for
the operators of the first-order Hamiltonian we get for scalar
operators
\beq{3.6}
\begin{array}{l}
\langle J_fM_f|{\ds\sum\limits_{j=1}^A}
e^{-i{\bf k}_{\nu}{\bf r}_j}
\left\{\begin{array}{c}
\hat 1\\
(\vsigma_j\hat{\bf p}_j)\\
iU'(r_j)(\vsigma_j{\bf n}_j)
\end{array}\right\}
\hat\tau_-(j)|J_iM_i\rangle={}
\\[\bigskipamount]
\phantom{\langle J_fM_f|{\ds\sum\limits_{j=1}^A}
e^{-i{\bf k}_{\nu}}}=
(4\pi)^{3/2}{\ds\sum\limits_{um}}(-1)^u
Y_{um}^*({\bf n}_{\nu})
C^{J_fM_f}_{J_iM_ium}
\left\{\begin{array}{c}
\mbox{$[0uu]$} \\
\mbox{$[0uu,p]$} \\
\mbox{$[0uu,r]$}
\end{array}\right\},
\end{array}
\eeq
and for $q$-th spherical components of vector operators
\beq{3.7}
\begin{array}{l}
\langle J_fM_f|{\ds\sum\limits_{j=1}^A}
e^{-i{\bf k}_{\nu}{\bf r}_j}
\left\{\begin{array}{c}
\sigma_{jq}\\
\hat p_{jq}\\
i\mbox{$[\vsigma_j\times\hat{\bf p}_j]$}_q\\
\mbox{$[[\vsigma_j\times\hat{\bf p}_j]\times\hat{\bf p}_j]$}_q\\
U'(r_j)\mbox{$[\vsigma_j\times {\bf n}_j]$}_q\\
\sigma_{jq}({\bf n}_{\nu}\hat{\bf p}_j)
\end{array}\right\}
\hat\tau_-(j)|J_iM_i\rangle={}
\\[\bigskipamount]
{}=\Frac{(4\pi)^{3/2}}{\sqrt{3}}{\ds\sum\limits_{wm}}(-1)^w
Y_{wm}^*({\bf n}_{\nu})
{\ds\sum\limits_{uM}}C^{uM}_{1qwm}C^{J_fM_f}_{J_iM_iuM}
\left\{\begin{array}{c}
\mbox{$[1wu]$}\\
\mbox{$[1wu,p]$}\\
\mbox{$[1wu,\sigma p]$}\\
\mbox{$[1wu,\sigma p^2]$}\\
\mbox{$[1wu,r]$}\\
\mbox{$\{1wu,\sigma p\}$}
\end{array}\right\}.
\end{array}
\eeq
The quantity $\{1wu,\sigma p\}$ is linear combination of the
reduced matrix elements
$[kwu,\sigma p]$ ($k=0,1$ or 2)
\beq{3.8}
\begin{array}{l}
\{1wu,\sigma p\}=(-1)^{w-u}{\ds\sum\limits_k}(-1)^k
\sqrt{\Frac{2k+1}{6(2u+1)}}\times{}
\\[\bigskipamount]
\phantom{\{1wu,\sigma p\}}\times
\left(\sqrt{w}\,U(w\,w\!-\!1\,1k,1u)[k\,w\!-\!1\,u,\sigma p]-
{}\right.
\\[\bigskipamount]
\left.\phantom{\{1wu,\sigma p\}=}-
\sqrt{w+1}\,U(w\,w\!+\!1\,1k,1u)[k\,w\!+\!1\,u,\sigma p]\right),
\end{array}
\eeq
defined by
\beq{3.9}
\begin{array}{l}
C^{J_fM_f}_{J_iM_iuM}[kwu,\sigma p]=
\sqrt{2}\,\langle J_jM_f|\sqrt{\Frac{3}{4\pi}}
{\ds\sum\limits_{j=1}^A}j_w(k_{\nu}r_j)\times{}
\\[\bigskipamount]
\phantom{C^{J_fM_f}_{J_iM_iuM}[kwu]}\times
{\ds\sum\limits_{nm}}
C^{uM}_{knwm}i^wY_{wm}({\bf r}_j)
\left({\ds\sum\limits_{\lambda q}}
C^{kn}_{1\lambda 1q}\sigma_{j\lambda}\hat p_{jq}\right)
\hat\tau_-(j)|J_iM_i\rangle.
\end{array}
\eeq
Note that $[0uu,\sigma p]=-\sqrt{2}[0uu,p]$. We use the normalized
Racah function $U(abcd,ef)=\sqrt{(2e+1)(2f+1)}\,W(abcd,ef)$. All
reduced matrix elements are real-valued quantities, provided that
the nuclear wave functions are transformed under time reversal in
standard way \cite{Boh69}
\beq{3.10}
\hat T|JM\rangle=(-1)^{J+M}|J\,-M\rangle.
\eeq

Let ${\bf n}_{\mu}$ be the unit vector along the axis $z$ of
orientation of the ensemble of the initial mesic atoms with given
spin $F$. Polarization and alignment of this ensemble are fixed by
spin-tensors $\tau_{10}(F)$ and $\tau_{20}(F)$ or normalized
parameters
\beq{3.11}
p_1(F)=\sqrt{\frac{F+1}{F}}\tau_{10}(F),\qquad
p_2(F)=\sqrt{\frac{(F+1)(2F+3)}{F(2F-1)}}\tau_{20}(F).
\eeq
Both parameters are equal unity if only the state corresponding to
the maximal projection $\xi=F$ is populated
($p_1(F)=\langle\xi\rangle/F$ is the conventional polarization).

After straightforward calculation we get from Eq.(\ref{3.5}) the
general expression for neutrino angular distribution
\beq{3.12}
\begin{array}{l}
\Frac{dw^F({\bf n}_{\nu})}{d\Omega}=
\Frac{C_{\mu}}{4\pi}\,
\Frac{2J_f+1}{2J_i+1}
{\ds\sum\limits_K}(2K+1)\tau_{K0}(F)
P_K({\bf n}_{\nu}{\bf n}_{\mu})\times{}
\\[\bigskipamount]
{}\times\left\{U(\Frac{1}{2}J_iFK,FJ_i)
{\ds\sum\limits_{uu'}}U(J_fu'J_iK,J_iu)
\left(C^{u'0}_{u0K0}V(u)V(u')-
\lefteqn{\phantom{\left\{\begin{array}{ccc}
w' & u' & 1\\
w  & u  & 1\\
N  & K  & 1
\end{array}\right\}}}\right.\right.
\\[\bigskipamount]
\phantom{\Frac{dw^F({\bf n}_{\nu})}{d\Omega}=
\Frac{C_{\mu}}{4\pi}\,
\Frac{2J_f+1}{2J_i+1}
{\ds\sum\limits_K}2K}-
\Frac{2}{\sqrt{3}}C^{u'0}_{u0K0}
\sum\limits_wC^{w0}_{u010}A(wu)V(u')+{}
\\[\bigskipamount]
\phantom{\Frac{dw^F({\bf n}_{\nu})}{d\Omega}}+
\Frac{1}{3}(-1)^{u'-u}{\ds\sum\limits_{ww'}}(-1)^{w'-w}
C^{w'0}_{w0K0}U(1w'uK,u'w)A(wu)A(w'u')+{}
\\[\bigskipamount]
{}+\Frac{\sqrt{2}}{3}{\ds\sum\limits_{ww'N}}
\sqrt{3(2N+1)(2u'+1)(2w+1)}
\left\{\begin{array}{ccc}
w' & u' & 1\\
w  & u  & 1\\
N  & K  & 1
\end{array}\right\}
C^{w'0}_{w0N0}C^{N0}_{10K0}\times{}
\\[\bigskipamount]
\left.\phantom{{}+\Frac{\sqrt{2}}{3}{\ds\sum\limits_{ww'N}}
\sqrt{3(2N+1)(2u'+1)(2w+1)}
\left\{\begin{array}{cc}
w' & u' \\
w  & u  \\
N  & K
\end{array}\right\}}\times
A(wu)A(w'u')\right)+{}
\\[\bigskipamount]
{}+{\ds\sum\limits_N}
\sqrt{6(2N+1)(2J_i+1)(2F+1)}
\left\{\begin{array}{ccc}
J_i & F & 1/2\\
J_i & F & 1/2\\
N   & K & 1
\end{array}\right\}
\sum\limits_{uu'}U(J_fu'J_iN,J_iu)\times{}
\\[\bigskipamount]
{}\times\left(
\lefteqn{\phantom{\left\{\begin{array}{ccc}
u  & w  & 1\\
u' & w' & 1\\
N  & K  & 1
\end{array}\right\}}}
C^{N0}_{K010}C^{u'0}_{u0N0}V(u)V(u')-
\Frac{2}{\sqrt{3}}{\ds\sum\limits_w}C^{u'0}_{w0K0}
U(u1u'K,wN)A(wu)V(u')-{}\right.
\\[\bigskipamount]
{}-2\sqrt{\Frac{2}{3}}
{\ds\sum\limits_w}A(wu)V(u')
{\ds\sum\limits_{\Lambda}}C^{u'0}_{w0\Lambda 0}
C^{\Lambda 0}_{K010}U(u1u'\Lambda,wN)
U(11NK,1\Lambda)-{}
\\[\bigskipamount]
{}-\Frac{1}{3}(-1)^{u'-u}C^{N0}_{K010}
{\ds\sum\limits_{ww'}}(-1)^{w'-w}
C^{w'0}_{w0N0}U(1w'uN,u'w)A(wu)A(w'u')+{}
\\[\bigskipamount]
{}+\Frac{2}{3}(-1)^N{\ds\sum\limits_{ww'}}
\sqrt{\Frac{2w'+1}{2w+1}}C^{u'0}_{w'010}
C^{w0}_{u'0K0}U(u1u'K,wN)A(wu)A(w'u')+{}
\\[\bigskipamount]
{}+\Frac{\sqrt{2}}{3}
{\ds\sum\limits_{ww'}}C^{w'0}_{w0K0}
\sqrt{3(2N+1)(2u'+1)(2w+1)}
\left\{\begin{array}{ccc}
u  & w  & 1\\
u' & w' & 1\\
N  & K  & 1
\end{array}\right\}\times{}
\\[\bigskipamount]
\left.\left.\phantom{{}+
{\ds\sum\limits_{ww'}}C^{w'0}_{w0K0}
\sqrt{3(2N+1)(2u'+1)(2w+1)}
\left\{\begin{array}{c}
u  \\
u' \\
N \end{array}\right\}}\times
A(wu)A(w'u')\right)\right\}.
\end{array}
\eeq
It is a series in the Legendre polynomials
\mbox{$P_0(\cos\theta)=1$}, $P_1(\cos\theta)=cos\theta$,
$P_2(\cos\theta)=3(\cos^2\theta-1)/2$\ldots, where $\theta$ is the
angle between ${\bf n}_{\nu}$ and ${\bf n}_{\mu}$. The constant is
\beq{3.13}
C_{\mu}=8(G\cos\theta_C)^2\left(\frac{mZe^2}{1+m/M_i}\right)^3
\frac{R(Z)E_{\nu}^2}{1+E_{\nu}/M_f},
\eeq
where $Z$ is the charge of the initial nucleus, and $R(Z)$ is the
correction factor for its non-pointlikeness. The amplitudes $V(u)$
and $A(wu)$ for the second-order Hamiltonian are given by
\beq{3.14}
V(u)=\left\{\begin{array}{ll}
G'_V[0uu]+G_1\Frac{\{1u,\sigma p\}}{M}+
G_2\Frac{\{1u,p\}}{M}, &
\mbox{if\enskip} \pi_i(-1)^u=\pi_f,\\
G'_P\{1u\}+g'_A\Frac{[0uu,p]}{M}+
g_P\Frac{[0uu,r]}{4M^2}, &
\mbox{if\enskip} \pi_i(-1)^u=-\pi_f,
\end{array}\right.
\eeq
\beq{3.15}
A(wu)=\left\{\begin{array}{ll}
G'_A[1wu]+g_2\Frac{[1wu,\sigma p^2]}{M^2}+
g_3\Frac{\{1wu,\sigma p\}}{M}, &
\mbox{if\enskip} \pi_i(-1)^w=\pi_f,\\
g'_V\Frac{[1wu,p]}{M}+g_1\Frac{[1wu,\sigma p]}{M}+
g_V\Frac{[1wu,r]}{4M^2}, &
\mbox{if\enskip} \pi_i(-1)^w=-\pi_f,
\end{array}\right.
\eeq
where
\beq{3.16}
\begin{array}{l}
\left\{\begin{array}{c}
\{1u\}\\
\{1u,p\}\\
\{1u,\sigma p\}\end{array}\right\}=
-\sqrt{\Frac{u}{3(2u+1)}}
\left\{\begin{array}{c}
\mbox{$[1\,u\!-\!1\,u]$}\\
\mbox{$[1\,u\!-\!1\,u,p]$}\\
\mbox{$[1\,u\!-\!1\,u,\sigma p]$}\end{array}\right\}+{}
\\[\bigskipamount]
\phantom{\left\{\begin{array}{c}
\{1u\}\end{array}\right\}=
-\sqrt{\Frac{u}{3(2u+1)}}
\left\{\begin{array}{c}
\mbox{999}\end{array}\right\}}+
\sqrt{\Frac{u+1}{3(2u+1)}}
\left\{\begin{array}{c}
\mbox{$[1\,u\!+\!1\,u]$}\\
\mbox{$[1\,u\!+\!1\,u,p]$}\\
\mbox{$[1\,u\!+\!1\,u,\sigma p]$}\end{array}\right\},
\end{array}
\eeq
$\pi_i$ and $\pi_f$ are parities of initial and final nuclear
states, respectively.

The case of transition $J_i\to J_f=J_i\pm 1$, $\pi_i=\pi_f$ with
dominating Gamow-Teller matrix element $[101]$ was considered in
Ref.\cite{Bar96}. The neutrino angular distribution has the form
\beq{3.17}
\begin{array}{l}
\Frac{dw^F({\bf n}_{\nu})}{d\Omega}=
\Frac{C_{\mu}[101]^2}{12\pi}
\Frac{2J_f+1}{2J_i+1}
\left(a_0+a_1p_1(F)({\bf n}_{\nu}{\bf n}_{\mu})+{}\right.
\\[\bigskipamount]
\left.\phantom{\Frac{dw^F({\bf n}_{\nu})}{d\Omega}=
\Frac{C_{\mu}[101]^2}{12\pi}
\Frac{2J_f+1}{2J_i+1}a_0+{}}+
a_2(F)p_2(F)\Frac{3({\bf n}_{\nu}{\bf n}_{\mu})^2-1}{2}\right),
\end{array}
\eeq
where
\beq{3.18}
a_0=G'_A(G'_A-\frac{2}{3}G'_P)C_1(J_i,J_f,F),
\eeq
\beq{3.19}
a_1=-G_A^{\prime 2}C_2(J_i,J_f,F)-
G'_A(G'_A-G'_P)C_3(J_i,J_f,F),
\eeq
\beq{3.20}
a_2=G'_AG'_PC_4(J_i,J_f,F).
\eeq
The explicit expressions for coefficients $C_i(J_i,J_f,F)$ are
presented in Ref.\cite{Bar96}.
Due to high sensitivity of the anisotropy $a_2$ to $G'_P$ the form
factor $g_P$ can be measured by this way. Note that this effect is
of first-order type ($G'_AG'_P\sim\ve$). Using the general
expression (\ref{3.12}) it is easy to estimate the contribution of
non-leading matrix elements for any given transition.

\section{Hyperfine effect}

The rate of muon capture from hyperfine state $|F_{\pm}\rangle$
($F_{\pm}=J_i\pm 1/2$) of mesic atom is given by isotropic term of
differential probability (\ref{3.12}), i.e.
\beq{4.1}
\Lambda_{\pm}={\ds\oint}\left(\frac{dw^{F_{\pm}}({\bf n}_{\nu})}
{d\Omega}\right)d\Omega.
\eeq
It can be represented in the form proposed in Ref.\cite{Wal76}
\beq{4.2}
\Lambda_F=\ov{\Lambda}+\delta\Lambda_F,
\eeq
where the statistically-averaged muon capture rate is given by
\beq{4.3}
\ov{\Lambda}=\frac{C_{\mu}}{2}\,
\frac{2J_f+1}{2J_i+1}\sum_u\left(x^2(u)+y^2(u)\right),
\eeq
and the hyperfine increment takes the form
\beq{4.4}
\begin{array}{l}
\delta\Lambda_F=\Frac{C_{\mu}}{2}\cdot
\Frac{2J_f+1}{2J_i+1}\cdot
\Frac{J_i(J_i+1)+3/4-F(F+1)}{2J_i(J_i+1)}\times{}
\\[\bigskipamount]
{}\times{\ds\sum\limits_u}\left(
\left(J_i(J_i+1)+u(u+1)-J_f(J_f+1)\right)
\left(\Frac{x^2(u)}{u}-\Frac{y^2(u)}{u+1}\right)-{}\right.
\\[\bigskipamount]
\phantom{\delta\Lambda}-
2\sqrt{(J_i\!+\!J_f\!+\!u\!+\!2)(J_i\!-\!J_f\!+\!u\!+\!1)
(J_i\!+\!J_f\!-\!u)(J_f\!-\!J_i\!+\!u\!+\!1)}\times{}
\\[\bigskipamount]
\left.\phantom{\phantom{\delta\Lambda_F}-
2\sqrt{(J_i\!+\!J_f\!+\!u\!+\!2)(J_i\!-\!J_f\!+\!u\!+\!1)
(J_i+{})}}\times
\Frac{x(u+1)y(u)}{u+1}\right),
\end{array}
\eeq
where
\beq{4.5}
x(u)=\sqrt{\frac{2u}{2u+1}}V(u)-
\sqrt{\frac{2(u+1)}{3(2u+1)}}A(uu)+
\sqrt{\frac{2}{3}}A(u-1\,u),
\eeq
\beq{4.6}
y(u)=\sqrt{\frac{2(u+1)}{2u+1}}V(u)+
\sqrt{\frac{2u}{3(2u+1)}}A(uu)-
\sqrt{\frac{2}{3}}A(u+1\,u).
\eeq
However, the muon capture rates $\Lambda_{\pm}$ may be brought
also to the form
\beq{4.7}
\begin{array}{l}
\Lambda_+=\Frac{C_{\mu}}{2}\,\Frac{2J_f+1}{2J_i+1}\,
\Frac{1}{2(J_i+1)}{\ds\sum\limits_u}\Frac{1}{u+1}\times{}
\\[\bigskipamount]
\phantom{\Lambda_+}\times\left(
\sqrt{(J_i+J_f-u)(J_f-J_i+u+1)}x(u+1)+{}\right.
\\[\bigskipamount]
\left.\phantom{\Lambda_+\times\sqrt{(J_i+J_f)}}+
\sqrt{(J_i+J_f+u+2)(J_i-J_f+u+1)}y(u)\right)^2,
\end{array}
\eeq
\beq{4.8}
\begin{array}{l}
\Lambda_-=\Frac{C_{\mu}}{2}\,\Frac{2J_f+1}{2J_i+1}\,
\Frac{1}{2J_i}{\ds\sum\limits_u}\Frac{1}{u+1}\times{}
\\[\bigskipamount]
\phantom{\Lambda_-=\Frac{C_{\mu}}{2}\,}\times\left(
\sqrt{(J_i+J_f+u+2)(J_i-J_f+u+1)}x(u+1)-{}\right.
\\[\bigskipamount]
\left.\phantom{\Lambda_+\times\sqrt{(J_i+J_f+u)}}-
\sqrt{(J_i+J_f-u)(J_f-J_i+u+1)}y(u)\right)^2.
\end{array}
\eeq
Putting
\beq{4.9}
x(u)=\left\{\begin{array}{ll}
M_u(u), & \mbox{if\enskip} \pi_i(-1)^u=\pi_f,\\
-M_u(-u), & \mbox{if\enskip} \pi_i(-1)^u=-\pi_f,
\end{array}\right.
\eeq
\beq{4.10}
y(u)=\left\{\begin{array}{ll}
M_u(-u-1), & \mbox{if\enskip} \pi_i(-1)^u=\pi_f,\\
M_u(u+1), & \mbox{if\enskip} \pi_i(-1)^u=-\pi_f.
\end{array}\right.
\eeq
we get the formulas \cite{Bal67} for muon capture rates from
different hyperfine states expressed in terms of the amplitudes
$M_u(k)$.

On the other hand, using Eqs.(\ref{3.14}), (\ref{3.15}) we obtain
by this way the explicit expressions for the amplitudes $M_u(k)$
with the accuracy up to the second-order terms in $1/M$
\beq{4.11}
\begin{array}{l}
M_u(u)=\sqrt{\Frac{2}{2u+1}}\left\{
\sqrt{u}\,G'_V[0uu]-
\sqrt{\Frac{u+1}{3}}\,G'_A[1uu]+{}\right.
\\[\bigskipamount]
\phantom{}+
\sqrt{\Frac{2u+1}{3}}
\left(g'_V-\Frac{u}{2u+1}G_2\right)
\Frac{[1u\!-\!1\,u,p]}{M}+
\sqrt\Frac{u(u+1)}{3(2u+1)}\,
G_2\,\Frac{[1u\!+\!1\,u,p]}{M}-{}
\\[\bigskipamount]
{}-\sqrt\Frac{2u+1}{3}
\left(\Frac{u}{2u+1}G_1-g_1-\Frac{u+1}{2(2u+1)}g_3\right)
\Frac{[1u\!-\!1\,u,\sigma p]}{M}+{}
\\[\bigskipamount]
\phantom{M_u(u)=\sqrt{\Frac{2}{2u+1}}
\left\{\sqrt{u}\,G'_V\right\}}+
\sqrt{\Frac{u(u+1)}{3(2u+1)}}
\left(G_1+\Frac{1}{2}g_3\right)\Frac{[1u\!+\!1\,u,\sigma p]}{M}+{}
\\[\bigskipamount]
{}+\Frac{1}{2}\sqrt{\Frac{(u-1)(u+1)}{3(2u+1)}}g_3
\Frac{[2u\!-\!1\,u,\sigma p]}{M}-
\Frac{1}{2}\sqrt{\Frac{(u+1)(u+2)}{3(2u+1)}}g_3
\Frac{[2u\!+\!1\,u,\sigma p]}{M}-{}
\\[\bigskipamount]
\left.\phantom{M_u(u)=\sqrt{\Frac{2}{2u+1}}}-
\sqrt{\Frac{u+1}{3}}\,g_2\,\Frac{[1uu,\sigma p^2]}{M^2}+
\sqrt{\Frac{2u+1}{3}}\,g_V\,
\Frac{[1u\!-\!1\,u,r]}{4M^2}\right\},
\end{array}
\eeq
\beq{4.12}
\begin{array}{l}
M_u(-u-1)=\sqrt{\Frac{2}{2u+1}}\left\{
\sqrt{u+1}\,G'_V[0uu]+
\sqrt{\Frac{u}{3}}\,G'_A[1uu]-{}\right.
\\[\bigskipamount]
\phantom{}-
\sqrt{\Frac{2u+1}{3}}
\left(g'_V-\Frac{u+1}{2u+1}G_2\right)
\Frac{[1u\!+\!1\,u,p]}{M}-
\sqrt\Frac{u(u+1)}{3(2u+1)}\,
G_2\,\Frac{[1u\!-\!1\,u,p]}{M}+{}
\\[\bigskipamount]
{}+\sqrt\Frac{2u+1}{3}
\left(\Frac{u+1}{2u+1}G_1-g_1-\Frac{u}{2(2u+1)}g_3\right)
\Frac{[1u\!+\!1\,u,\sigma p]}{M}-{}
\\[\bigskipamount]
\phantom{M_u(u)=\sqrt{\Frac{2}{2u+1}}
\left\{\sqrt{u}\,G'_V\right\}}-
\sqrt{\Frac{u(u+1)}{3(2u+1)}}
\left(G_1+\Frac{1}{2}g_3\right)\Frac{[1u\!-\!1\,u,\sigma p]}{M}-{}
\\[\bigskipamount]
{}-\Frac{1}{2}\sqrt{\Frac{u(u-1)}{3(2u+1)}}\,g_3\,
\Frac{[2u\!-\!1\,u,\sigma p]}{M}+
\Frac{1}{2}\sqrt{\Frac{u(u+2)}{3(2u+1)}}\,g_3\,
\Frac{[2u\!+\!1\,u,\sigma p]}{M}+{}
\\[\bigskipamount]
\left.\phantom{M_u(u)=\sqrt{\Frac{2}{2u+1}}}+
\sqrt{\Frac{u}{3}}\,g_2\,\Frac{[1uu,\sigma p^2]}{M^2}-
\sqrt{\Frac{2u+1}{3}}\,g_V\,
\Frac{[1u\!+\!1\,u,r]}{4M^2}\right\},
\end{array}
\eeq
\beq{4.13}
\begin{array}{l}
M_u(-u)=\sqrt{\Frac{2}{2u+1}}\left\{
-\sqrt{\Frac{2u+1}{3}}
\left(G'_A-\Frac{u}{2u+1}G'_P\right)[1u\!-\!1\,u]-{}\right.
\\[\bigskipamount]
\phantom{M_u(-u)={}}-
\sqrt{\Frac{u(u+1)}{3(2u+1)}}\,G'_P\,[1u\!+\!1\,u]-
\sqrt{u}\left(g'_A+\Frac{1}{3}g_3\right)
\Frac{[0uu,p]}{M}+{}
\\[\bigskipamount]
\phantom{M_u(-u)={}}+
\sqrt{\Frac{u+1}{3}}\,g'_V\,\Frac{[1uu,p]}{M}+
\sqrt{\Frac{u+1}{3}}\left(g_1+\Frac{1}{2}g_3\right)
\Frac{[1uu,\sigma p]}{M}+{}
\\[\bigskipamount]
{}+\sqrt{\Frac{(u-1)(2u+1)}{6(2u-1)}}g_3
\Frac{[2u\!-\!2\,u,\sigma p]}{M}-
\Frac{1}{6}\sqrt{\Frac{(u+1)(2u+3)}{2u-1}}g_3
\Frac{[2uu,\sigma p]}{M}-{}
\\[\bigskipamount]
\left.{}-\sqrt{\Frac{2u+1}{3}}\,g_2\,\Frac{[1u\!-\!1\,u,\sigma
p^2]}{M^2}-
\sqrt{u}\,g_P\,\Frac{[0uu,r]}{4M^2}+
\sqrt{\Frac{u+1}{3}}\,g_V\,\Frac{[1uu,r]}{4M^2}\right\},
\end{array}
\eeq
\beq{4.14}
\begin{array}{l}
M_u(u+1)=\sqrt{\Frac{2}{2u+1}}\left\{
-\sqrt{\Frac{2u+1}{3}}
\left(G'_A-\Frac{u+1}{2u+1}G'_P\right)[1u\!+\!1\,u]-{}\right.
\\[\bigskipamount]
\phantom{M_u(-u)}-
\sqrt{\Frac{u(u+1)}{3(2u+1)}}\,G'_P\,[1u\!-\!1\,u]+
\sqrt{u+1}\left(g'_A-\Frac{1}{3}g_3\right)
\Frac{[0uu,p]}{M}+{}
\\[\bigskipamount]
\phantom{M_u(-u)}+
\sqrt{\Frac{u}{3}}\,g'_V\,\Frac{[1uu,p]}{M}+
\sqrt{\Frac{u}{3}}\left(g_1+\Frac{1}{2}g_3\right)
\Frac{[1uu,\sigma p]}{M}-{}
\\[\bigskipamount]
{}-\sqrt{\Frac{(u+2)(2u+1)}{6(2u+3)}}\,g_3\,
\Frac{[2u\!+\!2\,u,\sigma p]}{M}+
\Frac{1}{6}\sqrt{\Frac{u(2u-1)}{2u+3}}\,g_3\,
\Frac{[2uu,\sigma p]}{M}-{}
\\[\bigskipamount]
\left.{}-\sqrt{\Frac{2u+1}{3}}\,g_2\,\Frac{[1u\!+\!1\,u,\sigma
p^2]}{M^2}+
\sqrt{u+1}\,g_P\,\Frac{[0uu,r]}{4M^2}+
\sqrt{\Frac{u}{3}}\,g_V\,\Frac{[1uu,r]}{4M^2}\right\}.
\end{array}
\eeq
We see that the second-order terms do not change the general
expressions (\ref{4.2})-(\ref{4.4}) or (\ref{4.7}), (\ref{4.8})
for muon capture rates from hyperfine sublevels but only are added
to the known zero-order and first-order terms in the amplitudes
$M_u(k)$.

To study the magnitude of interference between the zero-order and
second-order terms with respect to the squared first-order terms
let us consider as an example a transition $3/2\to 1/2$ between
the states with the same parity. The muon capture rates are given
by
\beq{4.15}
\Lambda_+=\frac{3C_{\mu}}{8}
\left(M^2_1(2)+\frac{2}{\sqrt{15}}M_1(2)M_2(2)+
\frac{1}{15}M^2_2(2)+\frac{16}{15}M^2_2(-3)\right),
\eeq
\beq{4.16}
\Lambda_-=\frac{2C_{\mu}}{3}
\left(M^2_1(-1)+\frac{1}{16}M^2_1(2)-
\frac{\sqrt{15}}{8}M_1(2)M_2(2)+\frac{15}{16}M^2_2(2)\right),
\eeq
The first-order term $\sim G'_P[101]$ may exceed in the amplitude
$M_1(2)$ both the zero-order term $\sim (G'_A-2G'_P/3)[121]$ and
the other first-order velocity terms due to the dominance of
Gamow--Teller matrix element $[101]$, on the one hand, and the
high absolute value of the form factor $g_P$, on the other hand.
Just in this case the squared first-order term
$\sim G^{\prime 2}_P$ appears in the numerator of the ratio
(\ref{1.1}). Then it is apparent that the significant influence on
the magnitude of the squared amplitude $M_1(2)$ is beyond the
capabilities of any second-order term in it, as
\beq{4.17}
\begin{array}{l}
M_1(2)=\sqrt{\Frac{2}{3}}\left\{
-\Frac{\sqrt{2}}{3}\,G'_P\,[101]-
\left(G'_A-\Frac{2}{3}G'_P\right)[121]+{}\right.
\\[\bigskipamount]
\phantom{M_1(2)=\sqrt{\Frac{2}{3}}}+
\sqrt{2}\left(g'_A-\Frac{1}{3}g_3\right)
\Frac{[011,p]}{M}+
\Frac{1}{\sqrt{3}}\,g'_V\,\Frac{[111,p]}{M}+{}
\\[\bigskipamount]
{}+\Frac{1}{\sqrt{3}}\left(g_1+\Frac{1}{2}g_3\right)
\Frac{[111,\sigma p]}{M}+
\Frac{1}{6\sqrt{5}}\,g_3\,\Frac{[211,\sigma p]}{M}-
\sqrt{\Frac{3}{10}}\,g_3\,\Frac{[231,\sigma p]}{M}-{}
\\[\bigskipamount]
\left.\phantom{M_1(2)=\sqrt{\Frac{2}{3}}}-
g_2\,\Frac{[121,\sigma p^2]}{M^2}+
\sqrt{2}\,g_P\,\Frac{[011,r]}{4M^2}+
\Frac{1}{\sqrt{3}}\,g_V\,\Frac{[111,r]}{4M^2}\right\}.
\end{array}
\eeq
It is clear that the contribution of second-order terms does not
exceed several percents. However, the terms associated with
nucleon-nucleus potential, which were never considered earlier,
deserve special attention.

\section{Summary}

The second-order in $1/M$ non-relativistic Hamiltonian of muon
capture is obtained. For the first time the terms associated with
nucleon-nucleus potential are considered. General treatment of the
hyperfine effect in muon capture rate is given with taking into
account of the second-order terms. In particular, the generalized
expressions for the usually used amplitudes $M_u(k)$ are
presented. It is shown that in the situation of the dominance of
the squared first-order terms the interference of zero-order and
second-order terms has no significant influence. However, the
second-order corrections may be of interest due to their
sensitivity to the nuclear potential.

General expression for neutrino (recoil nucleus) angular
distribution in muon capture by nucleus with non-zero spin is
proposed. In the approximation of dominating Gamow--Teller matrix
element we discuss the high sensitivity to $g_P$ of the term
related with alignment of the initial mesic atom. Such alignment
may be realized in capture of muons by oriented atoms. Using the
general formula for the angular distribution the contribution of
non-leading matrix elements to alignment effect can be easy
estimated for any given transition.
\bigskip\bigskip

I am grateful to Prof. J.P.Deutsch for his interest to this work
and useful discussions. The work was supported in part by grants
96-15-96548 and 96-02-17517 of Russian Foundation for Basic
Research.

\end{document}